\documentclass[preprint,12pt]{elsarticle}

\usepackage{amssymb}
\usepackage[none]{hyphenat}
\usepackage{enumerate,letltxmacro}
\usepackage{fixltx2e}
\usepackage{graphicx}
\LetLtxMacro\itemold\item
\renewcommand{\item}{\itemindent1.5em\itemold}

\journal{IEEE Frontiers in Education}

\begin{document}

\begin{frontmatter}



\title{Promoting Inclusive Design Practice at the Global Game Jam: A Pilot Evaluation}


\author{Michael James Scott, Gheorghita Ghinea \& Ian Hamilton}

\address{Brunel University London, UK}

\begin{abstract}
Games are a popular form of entertainment. However, many computer games present unnecessary barriers to players with sensory, motor and cognitive impairments. In order to overcome such pitfalls, an awareness of their impact and a willingness to apply inclusive design practice is often necessary. The Global Game Jam offers a potential avenue to promote inclusive design practices to students of game development. As such, this paper evaluates the impact of an initiative to promote inclusive design practices during the 2014 Global Game Jam. An attitude questionnaire was distributed to both participants and non-participants at one event venue. The results indicate that, having enrolled in the initiative, students' attitudes improved. Furthermore, all attendees reported they were likely to pursue further learning opportunities and consider accessibility issues in their future games. This suggests that the Global Game Jam, and other similar events, present an attractive avenue to promote inclusive design practice within the context of digital game development. However, further analysis of submitted games, additional qualitative inquiry and a large-scale trial are needed to determine impact on practice and to form recommendations for future events.
\end{abstract}


\end{frontmatter}


\section{Introduction}
Playing computer games is a popular form of entertainment, driving sales in excess of \$64.4 billion within the United States (US) between 2010 and 2013 \cite{1}. Among the factors for this success is the broad appeal of computer games. Research shows that computer games are enjoyed by a diverse audience. Approximately 71\% of players are adults and 48\% of players are female \cite{1}. Even members of groups not commonly associated with games, such as the elderly, sometimes report that they regularly play games \cite{2, 3}. Thus, as Allaire writes, ``there is no longer a `stereotype game player', but instead a game player could be your grandparent, your boss, or even your professor'' \cite{4}. However, different players often have different needs. It is, therefore, important to consider how this apparent emergence of diversity can be accommodated in order to maximize the market \cite{5}. 

One such community to consider are those with sensory, motor or cognitive impairments. Contrary to popular belief, the proportion of such players is not trivially small. More than one fifth of PopCap's casual games audience identified as having some form of impairment \cite{6}. Furthermore, an estimate based on the 2002 US Census data suggests that 9\% of the population can encounter a reduced play experience as a result of an impairment and 2\% of the population might be unable to play games as a result of an impairment \cite{7}. So, although this is likely to be fewer in practice as not everyone plays games, the proportion of population that does play is growing and this trend is likely to continue \cite{1,7}.

Unfortunately, many computer games present unnecessary barriers to those with impairments \cite{8, 9}. Consequently, such games limit their potential market as they are not accessible to those individuals whom want to purchase and play them. The term ``digital outcasts'' \cite{10, 11} has appeared in order to describe these individuals. That is, those being excluded by the gulf between the development of new innovations and their more inclusive variants.

It is interesting to note, however, that many of the barriers these outcasts encounter can be addressed. Many of the recommendations listed by the International Game Developers Association (IGDA) \cite{12} and the authors of Game Accessibility Guidelines \cite{13} are far from insurmountable; often, being small in scale and reasonably low-cost to implement. Particularly, when inclusivity is considered during the early stages of development. Given that such changes have the potential to significantly broaden a game's audience, investing in inclusive design practice could yield satisfactory returns. As such, many inclusive design practices are feasible for commercial organisations to engage with.

Despite this case, it is not clear whether many game developers have: an awareness of the impact of impairments; a knowledge of how to overcome pitfalls; or a willingness to even consider making inclusive games. With the increase of institutions offering educational programmes that focus on game design and development, students could be prepared with appropriate knowledge and skills prior to joining the industry. However, how can educators nurture the relevant attitudes in an effective manner? 

One approach, which has previously been piloted by the IGDA Special Interest Group on Game Accessibility \cite{14}, is to promote inclusive design at hackathons and other game making events such as the Global Game Jam. This is claimed to be an effective strategy \cite{15}, however as the initiative was only launched in 2012, no formal studies have been conducted to verify its impact.  This paper begins to address this gap.

\section{Global Game Jam}
The Global Game Jam (GGJ) is a two-day game making event which takes place simultaneously at multiple physical locations across the world. It was founded in 2009 and while many similar game festivals and hackathons existed prior to the first event, it was the first to organise multiple physical venues. It is also considered to be the largest event of its kind in the world. There were 488 locations across 72 countries involved in GGJ 2014, which had 4,290 games projects being submitted \cite{16}. Participants work in small groups to rapidly prototype video games based around a common theme and set of constraints, which are revealed at the beginning of the event. The brief time span aims to encourage creative thinking and experimentation. As such, a range of innovative and artistic games are often produced each year. These are available to download from the official GGJ website and it is standard practice at many sites for groups to present their game designs and game prototypes to an audience.

The event can be popular with undergraduate students, particularly those enrolled on game design and development programmes as it presents a range of opportunities for self-development and learning \cite{17}. It is not exclusively students who participate, therefore students have the opportunity to work alongside  industry practitioners and educators in the field. Furthermore, the experience can highlight individual strengths and weaknesses; as well as provide opportunities to work collaboratively with people from other disciplines.

A way in which the GGJ has distinguished itself from other similar events, is the flexibility it welcomes. Often, local site-specific constraints are embraced in order to explore key social or design issues. This presents opportunities for students to work alongside advocates as they engage with such constraints, forming a type of induction.  As such, this presents an exciting opportunity for students to become aware of key industry issues and experience common pitfalls.

Addressing the topic of game accessibility, an Accessibility Challenge has been offered at several GGJ venues since 2012. This has prompted participants to develop a range of accessible games, such as Super Space Snake shown below in Figure I. This initiative became better supported in 2014 when the GGJ offered optional design constraints focusing on game accessibility.

\begin{figure}[h!]
  \caption{Accessible Game Made During GGJ 2012 Bristol \cite{18}}
  \centering
    \includegraphics[width=\textwidth]{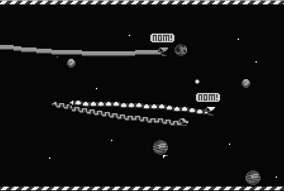}
\end{figure}

\section{Method}
The accessibility challenge was made available to students attending a GGJ 2014 location in London. This initiative took a minimalistic form, in which an advocate introduced the challenge and provided each group with handouts about visual, auditory, motor, and cognitive impairments as well as handouts on how to address each one. The advocate then periodically visited different groups throughout the event in order to encourage students to participate and to provide advice on their design ideas. 

In order to review the impact of the initiative, attitude questionnaires were distributed via SurveyMonkeyTM to everyone attending the event. This includes both those who opted-into the accessibility challenge and those who chose not to participate. No sampling was conducted as overall attendance was low. The following questions were posed:

\begin{table}[!t]
\renewcommand{\arraystretch}{1.3}
\caption{Game Accessibility Attitudes Questions}
\label{table:videos}
\centering
\begin{tabular}{p{2cm}p{10cm}}
\hline
    \textbf{Ref} & \textbf{Item} \\
\hline
    ATT-1 & I am familiar with the challenges that those with disabilities encounter when playing games \\
    ATT-2 & I consider making my game designs inclusive \\
    ATT-3 & I am aware of key game design pitfalls that affect how individuals with impairments enjoy games \\
    HARD-1 & I believe making an accessible game is too hard \\
    HARD-2 & I believe making an accessible game is too time consuming \\
\hline
\end{tabular}
\end{table}

These were presented as a 5-point Likert scale, from strongly disagree to strongly agree, and scores were computed by summation. Alongside demographic questions, several nominal questions were included in the post-event survey:

\begin{table}[!t]
\renewcommand{\arraystretch}{1.3}
\caption{Post-Event Impact Questions}
\label{table:videos}
\centering
\begin{tabular}{p{2cm}p{10cm}}
\hline
    \textbf{Ref} & \textbf{Item} \\
\hline
    IMPT-1 & Compared to everyone else, what kind of experience should disabled gamers expect to have? \\
    IMPT-2 & What level of priority should game developers adopt to avoid unnecessarily excluding gamers with disabilities? \\
    IMPT-3 & Would you like to meet other people who are interested in accessible game design? \\
    IMPT-4 & Would you be interested in taking part in a dedicated accessibility game jam? \\
\hline
\end{tabular}
\end{table}

In addition to several questions about participants' enjoyment of the event:

\begin{table}[!t]
\renewcommand{\arraystretch}{1.3}
\caption{Post-Event Enjoyment Questions}
\label{table:videos}
\centering
\begin{tabular}{p{2cm}p{10cm}}
\hline
    \textbf{Ref} & \textbf{Item} \\
\hline
ENJY-1 & The Global Game Jam is an enjoyable event \\
ENJY-2 & I believe attending the Global Game Jam is worthwhile \\
ENJY-3 & I would encourage others interested in games development to attend the Global Game Jam \\
\hline
\end{tabular}
\end{table}

These were presented as a 6-point forced-choice Likert scale, from strongly disagree to strongly agree, where scores were computed by summation.

\section{Results}

There were 35 complete responses to the survey, representing a response rate of 80\%. There were 11 first-year students, 10 second-year undergraduates, 2 final-year undergraduates, and 12 postgraduate students; of which, 5 were female and 23 had not previously attended a game jam. All cases were included and there was no missing data. All reported p-values from null hypothesis significance tests are two-tailed.

As participation in the accessibility challenge was optional, those who chose to participate (P; 10 respondents) and those who did not (NP; 25 respondents) have been analysed separately. Table IV below reveals a statistically significant improvement in ATT scores at post-test. Additionally, there was a statistically significant reduction in HARD scores for participants.

\begin{table}[!t]
\renewcommand{\arraystretch}{1.3}
\caption{Paired T-Test for Pre-Post Differences}
\label{table:videos}
\centering
\begin{tabular}{p{2cm}p{2cm}rrrrr}
\hline
\textbf{Items} & \textbf{Group} & \multicolumn{2}{l}{\textbf{Pre-Score}} & \multicolumn{2}{l}{\textbf{Post-Score}} & \textbf{$p$} \\
\textbf{} & \textbf{} & \textbf{$\mu$} & \textbf{$\sigma$} & \textbf{$\mu$} & \textbf{$\sigma$} & \textbf{} \\
\hline
    ATT   & P     & 9.7   & 3.3   & 11.8  & 3.01  & 0.003 \\
          & NP    & 9.76  & 2.9   & 11.1  & 3.23  & 0.013 \\
    HARD  & P     & 8.9   & 1.52  & 7.7   & 2.11  & 0.037 \\
          & NP    & 7.84  & 2.21  & 7.4   & 2.25  & 0.204 \\
\hline
\end{tabular}
\end{table}

Tables \ref{table:impt1}-\ref{table:impt4} below illustrate the IMPT responses for those attending the event:

\begin{table}[!t]
\renewcommand{\arraystretch}{1.3}
\caption{IMPT-1 Responses}
\label{table:impt1}
\centering
\begin{tabular}{p{8cm}rrr}
\hline
\textbf{Response} & \multicolumn{2}{r}{\textbf{Group}} & \textbf{Total} \\
\hline
    As equal as possible & 3     & 9     & 12 \\
    Roughly equivalent & 1     & 3     & 4 \\
    Access to some key areas of gameplay & 5     & 6     & 11 \\
    Separate disability-specific games & 1     & 7     & 8 \\
\hline
\end{tabular}
\end{table}

\begin{table}[!t]
\renewcommand{\arraystretch}{1.3}
\caption{IMPT-2 Responses}
\label{table:impt2}
\centering
\begin{tabular}{p{8cm}rrr}
\hline
\textbf{Response} & \multicolumn{2}{r}{\textbf{Group}} & \textbf{Total} \\
\hline
    Essential & 1     & 3     & 4 \\
    Only if it fits within the budget and mechanic & 8     & 16    & 24 \\
    Only after the game has been built & 1     & 4     & 5 \\
    Not at all & 0     & 2     & 2 \\
\hline
\end{tabular}
\end{table}

\begin{table}[!t]
\renewcommand{\arraystretch}{1.3}
\caption{IMPT-3 Responses}
\label{table:impt3}
\centering
\begin{tabular}{p{8cm}rrr}
\hline
\textbf{Response} & \multicolumn{2}{r}{\textbf{Group}} & \textbf{Total} \\
\hline
    Yes, I am interested in meeting other people & 9     & 10    & 19 \\
    No    & 1     & 15    & 16 \\
\hline
\end{tabular}
\end{table}

\begin{table}[!t]
\renewcommand{\arraystretch}{1.3}
\caption{IMPT-4 Responses}
\label{table:impt4}
\centering
\begin{tabular}{p{8cm}rrr}
\hline
\textbf{Response} & \multicolumn{2}{r}{\textbf{Group}} & \textbf{Total} \\
\hline
    Yes, I would be interested & 9     & 14    & 23 \\
    No    & 1     & 11    & 12 \\
\hline
\end{tabular}
\end{table}

These responses demonstrate a variety of perspectives with respect to each item. A series of chi-square tests show that there were no statistically significant differences in terms of response distribution for those who did and did not participate in the accessibility challenge. There was no general consensus regarding the experience that those with an impairment should expect. However, most of the respondents believed that inclusive designs should only be considered where they ``fit within the budget and intended game mechanics''.

Tables IX below shows the ENJY responses for those attending the event:

\begin{table}[!t]
\renewcommand{\arraystretch}{1.3}
\caption{Independent T-Test for Differences in Enjoyment}
\label{table:videos}
\centering
\begin{tabular}{p{2cm}p{4cm}rrr}
\hline
\textbf{Items} & \textbf{Group} & \textbf{$\mu$} &  \textbf{$\sigma$} & \textbf{$p$} \\
\hline
    ENJY  	& P     & 16.22 & 1.39  &  \\
      		& NP    & 16.28 & 1.69  & 0.928 \\
\hline
\end{tabular}
\end{table}

This shows that there was no statistically significant difference in the enjoyment of the Global Game Jam experience between those participating in the accessibility challenge and those choosing not to.

\section{Discussion}

The low rate of participation in the accessibility challenge was disappointing as other events have had higher rates of participation (e.g. \cite{19}). Further qualitative enquiry is needed to explore \textit{why} students did not want to engage with the challenge. Despite the low rate, however, the results suggest that the accessibility challenge encouraged all of the students to question their attitudes about game accessibility.

In particular, students noted increased awareness about the impact of impairments and clearly considered how they could overcome this impact. As such, many claimed that they would consider inclusive design practices in future projects. Therefore, the challenge appears to have had an impact. However, there was no consensus regarding the level of accessibility games should target, with many believing that games should be as 'equal as possible', with others believing that only 'access to key areas of gameplay' is sufficient.

It is important to note that only those whom participated in the accessibility challenge, began to dispel the belief that accessible games are difficult and time-consuming to create. As such, the game making experience does seem to have provided an attitude-changing experience (as claimed in \cite{15}). 

It should be noted that the sample size was small and all of the participants were drawn from a single venue. A larger trial conducted across multiple venues would permit a more representative and general conclusion.

There was only one form of measurement: a self-report questionnaire. Triangulation, through exploring students' efforts and analysing their future games, would permit a more sound interpretation of whether students' behaviours had actually changed as a result of their involvement in the event. As such, acquiescence bias poses a potential threat to the validity of these initial findings.

Finally, there was no formal assessment of whether the students were able to apply inclusive design practices effectively during such a short and intensive event. This was done in order to maintain a light-hearted and creative spirit. However, it makes it unclear whether the students learned how to apply inclusive design practice. It is proposed that formal observation at future events could be used to determine whether students adopted and applied appropriate practices. Additionally, further analysis of the games developed by the students prior to and after the event is needed to determine whether or not their experience transfers to new contexts as this is pertinent to the long-term impact of the initiative.

Depending on the results, it may be necessary  to consider the format of the event in greater detail. Particularly, how best practices can be facilitated. The nature of the event aims to encourage innovation and creativity, so a prescriptive approach may contrast with its culture. On the other hand, a trade-off with encouraging best practices may exist.

\section{Conclusion}

This paper shows that the Global Game Jam can be an effective avenue to promote inclusive design practice to students. A statistically significant improvement between pre-event and post-event attitudes were found with respect to knowledge about game accessibility and beliefs that making a game accessible is difficult. As such, the initiative appears to have increased students' awareness about game accessibility in addition to their willingness to consider accessibility issues in their future games. These are, hopefully, attitudes that the students will take with them when they become members of the games industry. 

Interestingly, despite low participation, non-participants showed some improvement in attitude. The reason for this change is not clear. Perhaps considering to participate, passing engagement with the material, or observing what other students achieved during the event were key factors. Nevertheless, the reasons for such low participation at this particular venue and the impact the challenge has on design behaviour (and the inclusivity of future games) warrant further investigation.

\section{Acknowledgement}

The authors would like to thank Chris Cox, and the School of Arts, who secured funding from Brunel University as well as helped to organise and run the event.

\section{References}



%\bibliographystyle{elsarticle-num} 
%\bibliography{compsoc}

\end{document}